%% file: Infrequent.tex
\documentclass{article}
\input{makros2.tex}

\usepackage{amsmath,amssymb,psfig}

\newcommand{\defeq}{\stackrel{\mathrm{def}}{=}}

\def\RR{\mathbb R}
\def\mI{\mathbb I}
\def\ZZ{\mathbb Z}

\def\mP{\mathbb P}

\def\cA{\mathcal A}
\def\cB{\mathcal B}
\def\cC{\mathcal C}
\def\cD{\mathcal D}
\def\cF{\mathcal F}
\def\cG{\mathcal G}

\def\cH{\mathcal H}
\def\cS{\mathcal S}

 \def\cP{\mathcal P}
\def\cL{\mathcal L}

\def\cQ{\mathcal Q}
\def\cR{\mathcal R}
\def\cU{\mathcal U}

\def\cH{\mathcal H}

\def\cX{\mathcal X}
\def\cY{\mathcal Y}

\newcommand{\fdtb}{\operatorname{TD}}
\newcommand{\pr}{\operatorname{PROJECT}}
\newcommand{\pd}{\operatorname{POLY-DUAL}}

\def\In{{\mathcal I}(\cA)}
\def\cI{{\mathcal I}}

\newcommand{\qed}{\hfill$\square$\bigskip}
\newcommand{\raf}[1]{(\ref{#1})}
\newcommand{\proof}{\noindent {\bf Proof}.\ }
\newcommand{\Min}{\operatorname{Min}}
\newcommand{\Max}{\operatorname{Max}}

\newcommand{\indeg}{\operatorname{d^{\bot}}}
\newcommand{\outdeg}{\operatorname{d^{\top}}}
\newcommand{\poly}{\operatorname{poly}}

\newcommand{\hide}[1]{}

\newtheorem{theorem}{Theorem}
\newtheorem{lemma}{Lemma}
\newtheorem{corollary}{Corollary}
\newtheorem{proposition}{Proposition}
\newtheorem{example}{Example}
\newtheorem{definition}{Definition}

\title{On Finding Minimal Infrequent Elements in Multi-dimensional Data Defined over Partially Ordered Sets}

\author{
Khaled M. Elbassioni
\thanks{
Masdar Institute of Science and Technology, P.O.Box 54224, Abu Dhabi, UAE;
(kelbassioni@masdar.ac.ae)}
}

\date{}

\begin{document}

\maketitle
\begin{abstract}
We consider databases in which each attribute takes values from a partially ordered set (poset). This allows one to model 
a number of interesting scenarios arising in different applications, including quantitative databases, taxonomies, 
and databases in which each attribute is an interval representing the duration of a certain event occurring over time.
A natural problem that arises in such circumstances is the following: given a database $\cD$ and a threshold value $t$,
find all collections of "generalizations" of attributes which are "supported" by less than $t$ transactions from $\cD$. 
We call such collections infrequent elements. Due to monotonicity, we can reduce the output size by considering only 
\emph{minimal} infrequent elements. We study the complexity of finding all minimal infrequent elements for some 
interesting classes of posets. We show how this problem can be applied to mining association rules 
in different types of databases, and to finding "sparse regions" or "holes" in quantitative data or in databases recording 
the time intervals during which a re-occurring event appears over time. Our main focus will be on these applications 
rather than on the correctness or analysis of the given algorithms. 

{\bf Keywords}: Association rules, categorical attributes, enumeration algorithms, frequent/infrequent elements, intervals, lattices, maximal empty boxes, partially ordered sets, quantitative data, rare associations, taxonomies.   
\end{abstract}

\section{Introduction}\label{s1}
The problem of mining association rules from large databases has emerged as an important area of research since their introduction in \cite{AIS93}. Typically, the different data
attributes exhibit certain correlations between them, which can be summarized in terms of certain rules, provided that enough transactions or records in the database agree with these rules.
For a few examples, in a database storing sets of items purchased by different customers in a supermarket, it may be interesting to observe a rule of the from "most customers that purchase 
bread and butter tend also to purchase orange juice"; in a database storing personal data about individuals, it may be interesting to observe that "most individuals who are married and with age in the range 28-34 have at least 2 cars"; and in a database storing data about the time periods a given service is used by different customers, an interesting observation may take the form: "     customers who make full use of the service between 2:00-3:00 on Friday tend also to use the service between 2:00-3:00 on Saturday".
Such information could be useful, for example, for placing items next to each other on supermarket shelves or providing better services for anticipated customers. 

\medskip

Most of the work on finding association rules divides the task into two basic steps: the first one is to identify those collections of items or attribute values that appear together frequently in the database, the so-called \emph{frequent} itemsets; the second step is to generate association rules from these. While the first step has received considerable attention in the literature, with many algorithms proposed, the second step seems to be somehow overlooked. In this chapter, we will have a more careful look at this latter step, and 
show in fact that a lot of redundancy can be eliminated from the generated rules by solving the somewhat complementary problem of finding \emph{infrequent sets}, i.e., those collection of items that \emph{rarely} appear together in any transaction. This gives one important motivation for studying the problem of finding infrequent collections of values that can be assumed by the attributes of the given database. But apart from that, finding such collections is a problem of independent interest, since each infrequent collection of attribute values indicates rare associations between these values. For instance, in the database of personal data above one can observe a rule like "no individuals with age between 26 and 38 have a single car", and in the database recording service usage, one may observe that ``Fewer than 40\% of the customers occupy the service on Friday between
2:00-3:00 and on Saturday between 2:00-4:00''. Another application will be given in Section \ref{rareAssoc}, in which the objective is to discover the so-called \emph{rare association rules},
which are informally rules that result from data appearing rarely in the database.  
  
\medskip

Rather than using \emph{binarization}, as is common in the literature (see e.g. \cite{SA95,SA96}), to represent the different ranges of each attribute by binary values, we shall consider more generally databases in which each attribute assumes values belonging to a \emph{partially ordered set} (poset). This general framework will allow us to model a number of different scenarios in data mining applications, including the mining of association rules for databases with quantitative, categorical and hierarchical attributes, and the discovery of missing 
associations or ``holes" in data (see \cite{AMSTV96,LKH97,LWMQ98}). One important feature of this framework is that it allows us to find \emph{generalized associations}, which are obtained by
generalizing some attribute values, for which otherwise there exist no enough support from the database transactions. As an example on the supermarket data above, it may be the case that most customers who purchase milk products tend also to purchase bread, but in the database only "cheese" and "butter" appear as items. In this case generalizing both these items to "milk products" allows us to discover the above rule.
 
\medskip
    
We begin our exposition in the next section with recalling some definitions and terminology related to partially ordered sets and give some examples of databases defined over products of posets. In Section \ref{s3}, we define the main object of interest in this chapter, namely minimal infrequent elements in products of posets, describe the associated enumeration problem, and discuss how to measure its complexity. Section \ref{s4} gives some applications of such enumeration problems to finding association rules in different types of databases and to finding sparse regions on quantitative data. In Section \ref{s5}, we discuss briefly the complexity of finding infrequent/minimal infrequent elements, with more details provided in the appendix for the interested reader. We conclude in Section \ref{s6} with pointers to implementation issues and some open problems.  

\section{Databases defined on products of partially ordered sets}\label{s2}
\begin{figure}[t]
  \centerline{
\parbox{3in}{\psfig{figure=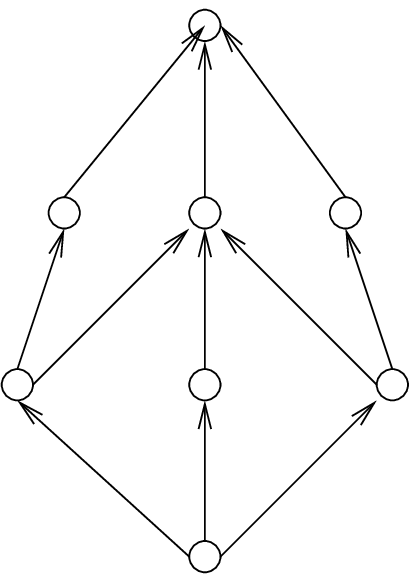,width=1.1in,height=1.0in}}
\hspace{0.5in}
\parbox{2.75in}{\psfig{figure=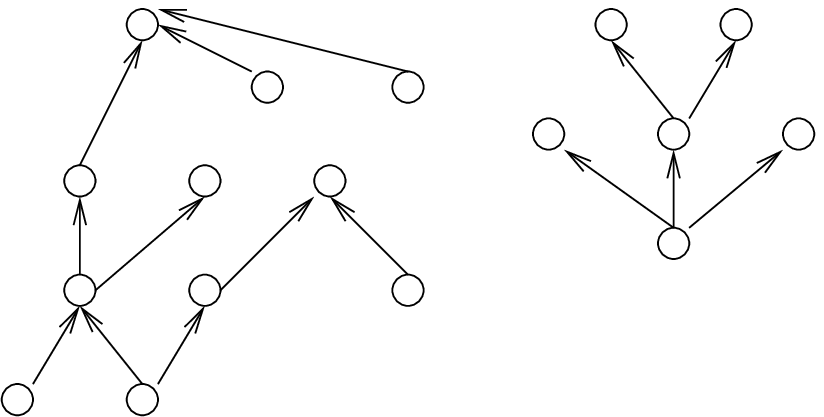,width=1.2in,height=1.0in}}
\hspace{0.5in}
\parbox{2.75in}{\psfig{figure=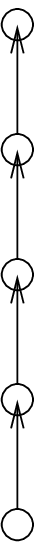,width=0.08in,height=1.0in}}}
\vspace{0.15in}
  \centerline{\hspace{0.6in} a: A lattice.  \hspace{1.0in} b: A forest. \hspace{0.5in} b: A chain. }
\caption{Lattices, forests and chains.}
\label{f1}
\end{figure}

Recall that a partially ordered set (poset) is defined by a pair $(\cP,\preceq)$, where $\cP$ is a finite set and $\preceq$ is a binary relation
satisfying the following three properties:
\begin{enumerate}
\item \emph{reflexivity}: $a\preceq a$ for all $a\in \cP$;
\item \emph{anti-symmetry}: if $a\preceq b$ and $b\preceq a$ then $a=b$;
\item \emph{transitivity}: if $a\preceq b$ and $b\preceq c$ then $a\preceq c$. 
\end{enumerate}
Let $\cP$ be (the ground set of) a poset. Two elements $x,y$ in $\cP$ are said to be \emph{comparable} if either $x\preceq y$ or $y\preceq x$ and otherwise are said to
be \emph{incomparable}. A \emph{chain} (anti-chain) of $\cP$ is subset of pairwise comparable (respectively, incomparable) elements.
For an element $x$ in $\cP$, we say that $y\in\cP$ is an \emph{immediate successor} of $x$ if $y\succ x$ and there is no $z\in\cP_i$ such that $y\succ z \succ x$. \emph{Immediate predecessors} of $x$ are defined similarly. 
The \emph{precedence graph} of a poset $\cP$ is a directed acyclic graph with vertex set $\cP$, and set of arcs $\{(x,y)~:~y\mbox{ is an immediate successor of }x\}$.
We say that poset $\cP$ is a \emph{forest} (or a \emph{tree}) if the underlying undirected graph of the precedence graph of $\cP$ is a forest (respectively, a tree).  
For two elements $x,y\in\cP$, $z$ is called an upper (lower) bound if $z\succeq x$ and $z\succeq y$ (respectively, $z\preceq x$ and $z\preceq y$). 
A \emph{join semi-lattice} (\emph{meet semi-lattice}) is a poset in which every two elements $x,y$ have a unique minimum upper-bound, called the \emph{join}, $x\vee y$ (respectively,
a unique maximum lower-bound, called the \emph{meet}, $x\wedge y$). A lattice is a poset which is both a join and a meet semi-lattice.  
For a poset $\cP$, the dual poset $\cP^*$ is the poset with the same set of elements as $\cP$, but such that $x\prec y$ in
$\cP^*$ whenever $x\succ y$ in $\cP$. 
The unique class of posets in the intersection of forests and lattices is the class of totally ordered sets, in which every two elements are comparable. Since the precedence graphs of such posets is a path, we shall refer also to them as chains. (See Figure \ref{f1} for an example.) For a good introduction to the theory of posets, we refer the reader to \cite{S03}.

\bigskip

Let $\cP\defeq\cP_1\times\cdots\times\cP_n$ be the Cartesian product of $n$
partially ordered sets. We will overload notation and denote by $\preceq$ the precedence 
relation in $\cP$ and also in $\cP_1,\ldots,\cP_n$, i.e., if
$p=(p_1,\ldots,p_n)\in\cP$ and $q=(q_1,\ldots,q_n)\in\cP$, then
$p\preceq q$ in $\cP$ if and only if $p_1\preceq q_1$ in
$\cP_1$, $p_2\preceq q_2$ in $\cP_2$,\ldots, and $p_n\preceq q_n$ in
$\cP_n$. 

We consider a database $\cD\subseteq \cP$ of transactions,
each of which is an $n$-dimensional vector of attribute values over $\cP$.
This gives a fairly general framework that allows us to model many interesting scenarios.
Let us look at some examples.

\subsection{Binary databases}
Perhaps, the simplest example is when the database is used to store transactions representing subsets of
items purchased by different customers in, say, a supermarket. Formally, we have a set $I$ of $n$ items, 
and each record in the database is a $0/1$-vector representing a subset of $I$. Thus, each factor poset
$P_i=\{0,1\}$ and the product $\cP$ is the Boolean cube $\{0,1\}^n$. Table \ref{tb1} shows an example of a binary database 
$\cD$.     

\begin{table}[htbp]
\centering\small
\begin{tabular}{|c|c|c|c|c|c|c|} \hline
$TID$ & Bread & Butter & Cheese & Milk &  Orange Juice & Yogurt\\ \hline
$T_1$ & 1 & 1 & 1 & 1 & 1 & 1\\ \hline
$T_2$ & 1 & 1 & 1 & 0 & 0 & 0\\ \hline
$T_3$ & 1 & 1 & 0 & 1 & 1 & 1\\ \hline
$T_4$ & 1 & 1 & 1 & 0 & 1 & 0\\ \hline
$T_5$ & 1 & 1 & 1 & 0 & 0 & 1\\ \hline
$T_6$ & 1 & 0 & 0 & 0 & 1 & 0\\ \hline
$T_7$ & 1 & 1 & 1 & 1 & 1 & 1\\ \hline
$T_8$ & 0 & 1 & 1 & 1 & 0 & 0\\ \hline
$T_9$ & 1 & 1 & 0 & 0 & 1 & 0\\ \hline
$T_{10}$ & 1 & 1 & 1 & 1 & 1 & 1\\ \hline
\end{tabular}
\\
\caption{Supermarket data}
\label{tb1}
\end{table}
\subsection{Quantitative databases}
This is the direct generalization of binary databases to the case when each attribute can assume integer or real values instead of being only binary.
In a typical database, most data attributes can be classified as
either categorical (e.g., zip code, make of car), or
quantitative (e.g., age, income). Categorical attributes assume only a fixed number of discrete values, but typically, there 
are no precedence relations between these. For instance,
there is no obvious way to order zip codes, and therefore, each such attribute $a_i$ assumes values from an antichain,
which can be equivalently represented by different binary attributes each corresponding to one value of $a_i$.  
Quantitative attributes, on the other hand, are real-valued attributes which are totally ordered, but for which 
there might not exist any bound. However, given a database of $m$
transactions, the number of different values that a given quantitative
attribute can take is at most $m$. As we shall see later, for our purposes, we
may assume without loss of generality that the different values of
each quantitative attribute $a_i$ are in one-to-one correspondence with some
totally ordered set (chain) $\cP_i$. Thus a database $\cD$ with Boolean,
categorical, and quantitative attributes can be represented as a subset of a
poset $\cP=\cP_1\times\ldots\times\cP_n$, where each poset $\cP_i$ is
a chain or an antichain. Table \ref{tb2} gives an example of a quantitative database\footnote{taken from \cite{SA96}}. 

\begin{table}[htbp]
\centering\small
\begin{tabular}{|c|c|c|c|} \hline
$ID$ & Age & Married & NumCars \\ \hline
$I_1$ & 23 & No & 1  \\ \hline
$I_2$ & 25 & Yes & 1  \\ \hline
$I_3$ & 29 & No & 0 \\ \hline
$I_4$ & 34 & Yes & 2  \\ \hline
$I_5$ & 38 & Yes & 2  \\ \hline
\end{tabular}
\\
\caption{Quantitative data}
\label{tb2}
\end{table}

\subsection{Taxonomies}
\begin{figure}[htbf]
\centerline{
\parbox{3in}{\psfig{figure=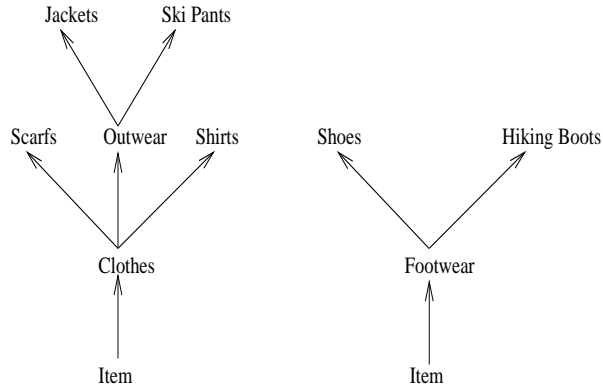,width=3.1in,height=2.0in}}
}
\caption{Example of a taxonomy}
\label{f2}
\end{figure}

This is yet another generalization of binary databases, in which each attribute can assume values belonging to some hierarchy.
For instance, in a store, items available for purchase can be classified into different categories, e.g., clothes, footwear, etc.
Each such type can be further classified, e.g., clothes into scarfs, shirts, etc. Then further classification are possible, and so on. 
Figure \ref{f2} gives an example of two such taxonomies. Typically, a database of transactions $\cD$ is given where each transaction represents the set of 
items purchased by some customer. Each such item is a top-level element in a certain hierarchy (e.g., scarfs, jackets, ski pants,
shirts, shoes, and hiking boots, in Figure \ref{f2}). To obtain generalized association rules which have enough support from the database, 
it may be necessary to generalize some items as described by the hierarchy (more on this in Section \ref{sgen}). This can be done by having each attribute
$a_i$ in the database assume values belonging to a tree poset $\cP_i$. To account for transactions that do not contain any element from a certain taxonomy, a minimum element called "Item"
is assumed to be at the lowest level in each taxonomy. For instance, in Table \ref{tb3}, transaction $T_6$ corresponds to the element $(\mbox{Jacket,Item})$. Then $\cD\subseteq\cP=\cP_1\times\cdots\times\cP_n,$ where $n$ is the number of different attributes. 
Table \ref{tb3} shows an example\footnote{taken from \cite{SA95}}, where
$n=2$ and the two posets $\cP_1$ and $\cP_2$ correspond to the two taxonomies shown in Figure \ref{f2}.

\begin{table}[htbp]
\centering\small
\begin{tabular}{|c|c|c|c|c||c|c|} \hline
& \multicolumn{4}{|c||}{Clothes} &  \multicolumn{2}{|c|}{Footwear} \\\cline{2-4}  \hline\hline
$TID$ & Jacket & Scarf & Shirt & Ski Pants & Hiking Boots & Shoes  \\ \hline
$T_1$ & 0 & 0 & 1 & 0 & 0 & 0 \\ \hline
$T_2$ & 1 & 0 & 0 & 0 & 1 & 0 \\ \hline
$T_3$ & 0 & 0 & 0 & 1 & 1 & 0 \\ \hline
$T_4$ & 0 & 0 & 0 & 0 & 0 & 1 \\ \hline
$T_5$ & 0 & 0 & 0 & 0 & 0 & 1 \\ \hline
$T_6$ & 1 & 0 & 0 & 0 & 0 & 0 \\ \hline
\end{tabular}
\\
\caption{A hierarchical database}
\label{tb3}
\end{table}

\subsection{Databases of events occurring over time}\label{s-time}
Consider the situation when each attribute in the database can assume an interval of time. 
For instance, a service provider may keep a log file containing the start and end times at which each customer has used the service\footnote{A more specific example, given in \cite{L03}, is a cellular phone company which records the time and length for each phone call made by each
customer.}. 
 To analyze the correlation between the usage of the service at different points of time, one discretizes the time horizon into 
$n$ regions. Naturally, these could be the days of the week ($n=7$) or the days of the year ($n=365$). For each such region, we get a collection of intervals 
$\mI_i$, $i=1,\ldots,n$, which represent the usage of the service during that region of time. We shall need the following definition.

\begin{definition}\label{d2}\emph{(Lattice of intervals)}
Let $\mI$ be a set of real closed intervals. The {\em lattice of
intervals} $\cP$ defined by $\mI$ is the lattice whose elements are all possible intersections and spans defined by the intervals in $\mI$, and ordered by
containment. The meet of any two intervals in $\cP$ is their {\it intersection}, and the join is their {\it span}, i.e., the
minimum interval containing both of them.
\end{definition}

\begin{figure}[t]
  \centerline{
\parbox{2.75in}{\psfig{figure=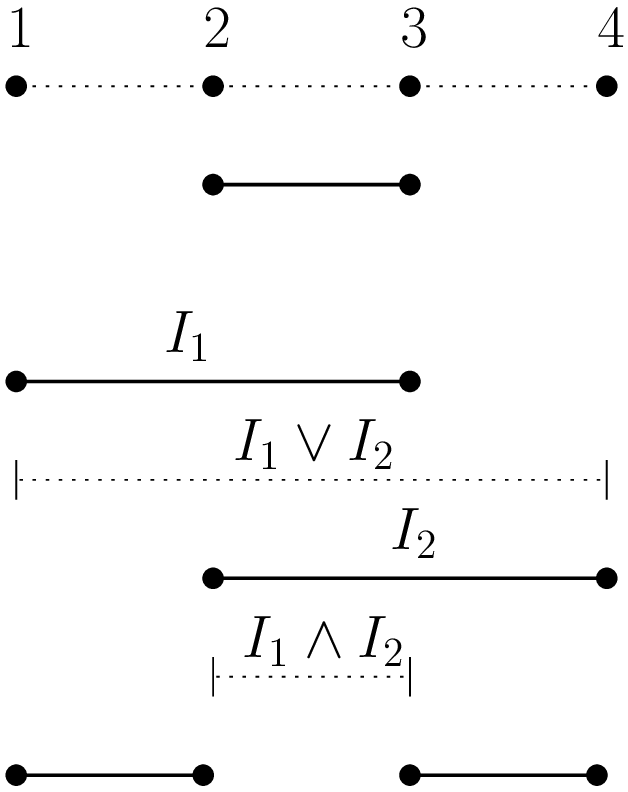,width=1.60in,height=1.5in}}
        \hspace{0.5in}
\parbox{3in}{\psfig{figure=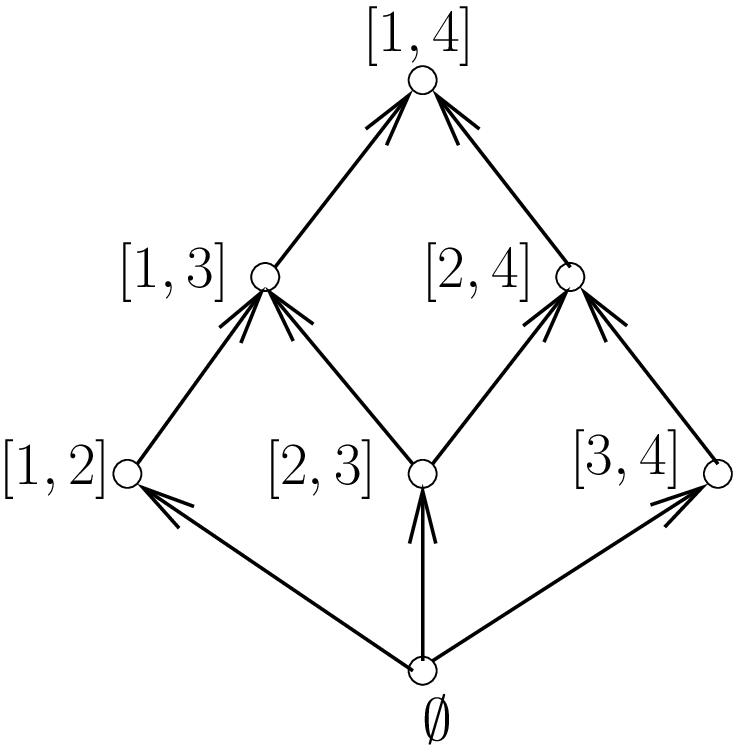,width=2.00in,height=2.1in}}}
\vspace{0.15in}
  \centerline{a: A set of intervals $\mI_1$. \hspace{0.5in} b: The
corresponding lattice of intervals $\cP_1$.}
\caption{The lattice of intervals.}
\label{f3}
\end{figure}

Consider for instance the database shown in Table \ref{tb4}. It shows the times of 3 days of the week at which a set of customers have visited a certain web server.  
Figure \ref{f3} gives the set of intervals defined by the first column of the database, and the corresponding lattice of intervals defined by them.

For $i=1,\ldots,n$, let $\cP_i$ be the lattice of
intervals defined by the intervals in $\mI_i$. Then we arrive at a scenario where the database $\cD$ is a subset of the lattice product $\cP=\cP_1\times\cdots\times\cP_n$.       

\hide{
\begin{table}[htbp]
\centering\small
\begin{tabular}{|c|c|c|c|c|c|c|c|} \hline
$TID$ & Monday & Tuesday & Wednesday & Thursday & Friday & Saturday & Sunday\\ \hline
$T_1$ & 5:30-6:00 & - & 3:00-6:00 & - & 9:00-12:00 & - & -  \\ \hline
$T_2$ & 5:15-7:00 & 9:00-11:21 & - & - & - & - & - \\ \hline
$T_3$ & 5:50-8:00 & 8:00-11:00 & - & - & 7:00-11:15 & - &  13:00-14:00 \\ \hline
$T_4$ & 7:30-10:45 & - & - & 14:00-18:00 & 7:00-8:00 & 9:00-11:00 &  9:00-11:00 \\ \hline
$T_5$ & 8:12-10:30 & 10:00-12:00 & 10:30-11:30 & 5:30-8:00 & - & 9:00-12:00 &  - \\ \hline
$T_6$ & 5:00-11:00 & - & 10:30-12:20 & 8:30-9:50 & - & 9:00-12:00 &  10:00-11:30 \\ \hline
\end{tabular}
\\
\caption{A database of intervals}
\label{table4}
\end{table}
}

\begin{table}[htbp]
\centering\small
\begin{tabular}{|c|c|c|c|} \hline
$TID$ & Friday & Saturday & Sunday\\ \hline
$T_1$ & 2:00-3:00 & 2:00-3:00 & 1:00-2:00  \\ \hline
$T_2$ & 1:00-3:00 & 1:00-3:00 & 1:00-3:00 \\ \hline
$T_3$ & 2:00-4:00 &  2:00-4:00 & 1:00-4:00\\ \hline
$T_4$ & 1:00-2:00 &  1:00-4:00 & - \\ \hline
$T_5$ & 3:00-4:00 &  -  & 1:00-3:00\\ \hline
\end{tabular}
\\
\caption{A database of intervals: "-" indicates no usage of the service}
\label{tb4}
\end{table}

\section{Infrequent elements}\label{s3}
\subsection{Definitions and notation}
In the following sections, we let $\cP=\cP_1\times\cdots\times\cP_n$ be a product of $n$ posets and  $\cD\subseteq\cP$ be a database defined over $\cP$.

\begin{definition}\label{d1}\emph{(Support)}
For an element $p\in\cP$, let us denote by
$$
  S(p)=S_{\cD}(p)\defeq \{q\in\cD~|~q\succeq p\},
$$
the set of transactions in $\cD$ that {\em support} $p\in\cP$.
\end{definition}

Note that the function $|S_{\cD}(p)|$ is \emph{monotonically non-decreasing} in $p\in\cP$, i.e., if $p\preceq q$, then $|S_{\cD}(p)|\geq |S_{\cD}(q)|$.

\begin{definition}\label{d3}\emph{(Frequent/infrequent element)}
Given $\cD\subseteq \cP$ and an integer threshold $t$, let
us say that an element $p\in\cP$ is $t$-frequent if it is supported by
at least $t$ transactions in the database, i.e., if $|S_{\cD}(p)|\geq
t$. Conversely, $p\in\cP$ is said to be $t$-infrequent if $|S_{\cD}(p)|<
t$. 
\end{definition}

Note that the property of being infrequent is \emph{monotone}, i.e., if $x$ is $t$-infrequent and $y\succeq x$, then $y$ is also $t$-infrequent. This motivates the following definition.

\begin{definition}\label{d4}\emph{(Minimal infrequent/maximal frequent element)}
An element $p\in\cP$ is said to be minimal $t$-infrequent (maximal $t$-frequent) with respect to a database $\cD\subseteq \cP$ and an integer threshold $t$, if $p$ is $t$-infrequent (respectively, $t$-frequent),
but any $q\in\cP$ such that $q\prec p$ (respectively, $q\succ p$) is $t$-frequent (respectively, $t$-infrequent).
\end{definition}

\begin{example}\label{e1}
Consider the binary database in Table \ref{tb1}. The set of items $X=\{\mbox{Bread,Butter}\}$ has support $|S(X)|=8$. For $t=4$, $X$ is $t$-frequent but not maximal as it is contained in the maximal $t$-frequent set $\{\mbox{Bread,Butter,Cheese,Orange Juice}\}$. The set  $\{\mbox{Bread,Butter,Cheese,Milk,Orange Juice, Yogurt}\}$ is $t$-infrequent but not minimal since
it contains the minimal $t$-infrequent set $\{\mbox{Bread,Butter,Cheese,Milk,Orange Juice}\}$.
\end{example}

\begin{example}\label{e2}
Consider the database in Table \ref{tb3}. The element $x=(\mbox{Outwear,Footwear})$ has support $|S(x)|=2$. For $t=2$, $x$ is $t$-frequent but not maximal as it precedes the maximal $t$-frequent element $(\mbox{Outwear,Hiking Boots})$. The element $(\mbox{Jacket, Hiking Boots})$ is $t$-infrequent but not minimal since
it is above the minimal $t$-infrequent element $(\mbox{Jacket,Footwear})$.
\end{example}

Given a poset $\cP$, and a subset of its elements $\cA \subseteq \cP$,
we will denote by $\cA^+=\{x \in \cP~|~x \succeq a, \mbox{ for some } a \in \cA\}$ and
$\cA^{-}=\{x \in \cP~|~x \preceq  a,\mbox{ for some } a \in  \cA\}$, the so-called \emph{ideal} and \emph{filter} defined by $\cA$. 

\begin{definition}\label{d5}\emph{(independent/maximal independent element)}
Let $\cP$ be a poset and $\cA$ be an arbitrary subset of $\cP$. An element in $p\in \cP$ is called independent of $\cA$ if 
$p$ is not above any element of $\cA$, i.e., $p\not\in\cA^+.$ $p$ is said further to be a maximal independent element if there is no
$q\in\cP$, such that $q\succ p$ and $q$ is independent of $\cA$.  
\end{definition}

Throughout we will denote by $\cI(\cA)$ be the set of all maximal independent elements for
$\cA$. Then one can easily verify the following decomposition of $\cP$:
\begin{equation}\label{eb}
\cA^+ \cap \In^{-} = \emptyset,~~~~\cA^+ \cup \In^{-}=\cP.
\end{equation}

Given a database $\cD\subseteq \cP$, and an integer threshold $t$, let
us denote by $\cF_{\cD,t}$ the set of minimal $t$-infrequent elements of $\cP$
with respect to $\cD$ and $t$. 

Then $\cI(\cF_{\cD,t})$ is the
set of maximal $t$-frequent elements:
$$
\cF_{\cD,t}=\Min\{x\in\cP~:~|S_{\cD}(x)|<
t\},~~~~\cI(\cF_{\cD,t})=\Max\{x\in\cP~:~|S_{\cD}(x)|\ge t\},
$$
where for a set $\cA\subseteq\cP$, we denote by $\Min(\cA)$
(respectively, $\Max(\cA)$), the smallest cardinality (with respect to the relation $\preceq$) set $\cB\subseteq\cP$ such that $\cB^+=\cA^+$
(respectively, $\cB^-=\cA^-$). Using the above notation, the sets $\cF_{\cD,t}^+$ and $\cI(\cF_{\cD,t})^-$ will denote respectively the
set of $t$-infrequent and $t$-frequent elements.

\subsection{Associated enumeration problems}
The problem of finding all frequent/infrequent elements in a database has proved useful in data mining applications \cite{GMKT97} (see also the examples below). 
As mentioned earlier, the property of being infrequent is monotone, and hence a lot of redundancy can be removed by considering only minimal $t$-infrequent elements.
This motivates us to study the complexity of the problem finding the sets $\cF_{\cD,t}$ and $\cI(\cF_{\cD,t})^-$ of all minimal $t$-infrequent elements and all $t$-frequent elements, respectively.
The generic generation problem we will consider is the following:

\begin{description}
\item[{\em GEN}$_\cH(\cP,\cD,t)$:] {\em Given a database $\cD$ defined over in a poset product $\cP$,
and a threshold $t$, find all elements of $\cH$ with respect to $\cD$ and $t$}.
\end{description}

In the above definition if $\cH=\cF_{\cD,t}$ then we are considering the generation of minimal infrequent elements, and if $\cH=\cI(\cF_{\cD,t})^-$ ($\cH=\cF_{\cD,t}^+$) then we are considering the generation of frequent (respectively, infrequent) elements. Clearly, the whole set $\cH$ can be generated by starting with $\cX=\emptyset$ and performing $|\cH|+1$ calls to the following incremental generation problem (with $k=1$):
 
\begin{description}
\item[{\em INC-GEN}$_\cH(\cP,\cD,t,\cX,k)$:] {\em Given a database $\cD$ defined over a poset product $\cP$, a threshold $t$, a subset $\cX\subseteq\cH$, and
an integer $k$, find $\min\{k,|\cH\setminus\cX|\}$ elements of $\cH\setminus
\cX$, or state that no such element exists}.
\end{description}

Before we talk about the complexity of the enumeration problems we are interested in, we should remark on how to measure this complexity, since typically the 
complete output size is exponentially large in the size of the input database. One can distinguish different notions
of efficiency, according to the time/space complexity of such generation problem:

\begin{itemize}
\item \emph{Output polynomial} or \emph{Total polynomial}:
Problem GEN$_{\cH}(\cP,\cD,t)$ can be
solved in $\poly(\sum_{i=1}^n|\cP_i|,|\cD|)$ time.
\item \emph{Incremental polynomial}: Problem INC-GEN$_{\cH}(\cP,\cD,t,\cX,1)$
can be solved in $\poly(\sum_{i=1}^n|\cP_i|,|\cD|,|\cX|)$ time, for every
$\cX\subseteq\cH$, or equivalently, INC-GEN$_{\cH}(\cP,\cD,t,\emptyset,k)$
can be solved in $\poly(\sum_{i=1}^n|\cP_i|,|\cD|,\min\{k,|\cH|\})$ time, for every integer $k$.
\item \emph{Polynomial delay}: INC-GEN$_{\cH}(\cP,\cD,t,\cX,1)$
can be solved in $\poly(\sum_{i=1}^n|\cP_i|,|\cD|)$ time.
In other words, the time required to generate
a new element of $\cH$ is polynomial only in the input size. If the time required to solve
INC-GEN$_{\cH}(\cP,\cD,t,\cX,1)$ is $\poly(\sum_{i=1}^n|\cP_i|,|\cD|)|\cX|$, then the problem is said to be solvable with
\emph{amortized} polynomial delay.
\item \emph{Polynomial space}:
The total space required to solve GEN$_{\cH}(\cP,\cD,t)$
is bounded by a $\poly(\sum_{i=1}^n|\cP_i|,|\cD|)$.
This is only possible if the algorithm looks at no more than
$\poly(\sum_{i=1}^n|\cP_i|,|\cD|)$ many outputs that it has already generated.
\item \emph{{\rm NP}-hard}: the decision problem associated with INC-GEN$_{\cH}(\cP,\cD,t,\cX,1)$ (i.e., deciding if $\cH=\cX$)
is NP-hard, which means that  is
coNP-complete, since it belongs to coNP.
\end{itemize}

\medskip

We will see that, generally, the generation of infrequent elements can be done with amortized polynomial delay, using Apriori-like algorithm,
while the currently best known algorithm for generating minimal infrequent elements runs in quasi-polynomial time.

The general framework suggested in this section allows us to model a number of different scenarios in data mining applications. We consider some examples in the next section.

\section{Applications}\label{s4}
\subsection{Mining association rules}
\subsubsection{Boolean association rules}
Consider a binary database $\cD$ each record of which represents a subset of items from a large set $V$ of $n$ items.  
In our terminology, we have $\cP_i=\{0,1\}$ for $i=1,\ldots,n$, and $\cD\subseteq\cP=2^{V}$, the binary cube of dimension
$n$. We recall the following central definition from \cite{AIS93}:

\begin{definition}\label{d6}\emph{(Association rules)}
Let $\cD\subseteq 2^V$ be a binary database, and $s,c\in[0,1]$ be given numbers. 
An association rule, with support $s$ and confidence $c$, is a pair of disjoint subsets 
$X,Y\subseteq [n]$
such that
$$
\frac{|S_{\cD}(X\cup Y)|}{|S_{\cD}(X)|}\geq c,~~~~~~~\frac{|S_{\cD}(X\cup Y)|}{|\cD|} \geq s,
$$
and will abbreviated by $X\Rightarrow Y |(c,s)$. (That is, at least $c$ fraction of the transactions that contain $X$ also contain $Y$ (confidence condition), and
at least a fraction $s$ of all transactions contain both $X$ and $Y$ (support condition).)  

\end{definition}

Each such rule $X\Rightarrow Y$ roughly means that transactions which
contain all items in $X$ tend also to contain all items in $Y$. Here
$X$ is usually called the {\em antecedent} of the rule, and $Y$ is called
the {\em consequent}. Generating such association rules has received a lot of attention
since their introduction in \cite{AIS93}. 

Note that the anti-monotonicity of the support function
implies the following. 
\begin{proposition}\label{p1}
Let $X,Y,X',Y'\subseteq V$ be such that $X'\supseteq X$ and $X'\cup Y'\subseteq X\cup Y$, and suppose that the rule $X\Rightarrow Y |(c,s)$ holds.
Then the rule $X'\Rightarrow Y' |(c,s)$ also holds. 
\end{proposition}
\proof
Set $Z=X\cup Y$ and $Z'=X'\cup Y'.$ Then $|S_{\cD}(Z)|\geq s|\cD|$ and $|S_{\cD}(X)|\leq |S_{\cD}(Z)|/c$ since the rule $X\Rightarrow Y |(c,s)$ holds.
Since $X'\supseteq X$ and $Z'\subseteq Z$, we get
\begin{eqnarray*}
|S_{\cD}(Z')|&\geq& |S_{\cD}(Z)|\geq s|\cD|\\
|S_{\cD}(X')|&\leq& |S_{\cD}(X)|\leq \frac{|S_{\cD}(Z)|}{c}\leq\frac{|S_{\cD}(Z')|}{c}.
\end{eqnarray*}
\qed

Clearly, one should be interested only in generating rules that are not implied
by others. 
This motivates the following definition.

\begin{definition}\label{d7}\emph{(Irredundant association rules)}
Let $\cD\subseteq 2^V$ be a binary database, and $s,c\in[0,1]$ be given numbers. 
An irredundant association rule  $X\Rightarrow (Z\setminus X) |(c,s)$, with support $s$ and confidence $c$, 
is determined by a pair of a (inclusion-wise) minimal subset 
$X$ and a maximal subset $Z$, such that $X\subseteq Z$, and
\begin{eqnarray}
\label{eq1}
|S_{\cD}(Z)|&\geq& s|\cD|\\
\label{eq2}
|S_{\cD}(X)|&\leq& \frac{|S_{\cD}(Z)|}{c}.
\end{eqnarray}
\end{definition}

\begin{example}\label{e3}
Consider the binary database in Table \ref{t1}. Using $s=0.4$ and $c=0.5$, one can verify that the rule $\{\mbox{Bread, Butter, Cheese}\}\Rightarrow\{\mbox{Orange Juice}\}$ holds. However,
this a redundant rule since it is implied by the irredundant rule $\{\mbox{Bread, Butter}\}\Rightarrow\{\mbox{Cheese, Orange Juice}\}$. 
\end{example}

It follows from Definition \ref{d7} that, in order to generate irredundant association rules, one needs to perform two basic steps (see Figure \ref{f-p1}):
\begin{enumerate}
\item Generate all subsets $Z$ satisfying \raf{eq1}; these are the elements of the family $\cI(\cF_{\cD,t})^-$ ($t$-frequent sets) where $t=s|\cD|$, which can be obtained by
solving problem GEN$_{\cI(\cF_{\cD,t})^-}(2^V,\cD,t)$. This can be done using the \emph{Apriori} algorithm; see Section \ref{s5} and Appendix A.
\item For each such $t$-frequent set $Z$, generate all minimal $t'$-infrequent subsets of $Z$, where $t'=|S_{\cD}(Z)|/c+1$. To avoid generating redundant rules, we maintain a list $\cX(Z)$ of already generated $t'$-infrequent subsets of $Z$. For each set $Z,$ we compute the set $\cX(Z)$ by solving problem GEN-INC$_{\cF_{\cD[Z],t'}}(2^{Z},\cD[Z],t',\cX',|\cF_{\cD[Z],t'}\setminus\cX'|)$, where $\cD[Z]=\{T\cap Z:~T\in \cD\}$, and $\cX'$ is the set of minimal infrequent subsets of $Z$ that are contained in some $X\in\cX(Z')$ for some $Z'\supseteq Z$. The set $\cX'$ can be computed easily once we have computed $\cX(Z')$ for all $Z'\supset Z$, and in particular all subsets $Z'$ that have one more item 
than $Z$. That is why the procedure iterates from larger frequent sets to small ones.    
\end{enumerate}

\begin{figure}
\begin{code}
\noindent{\bf Procedure GEN-RULES$(\cD,c,s)$:}\\
\> Input: A binary database $\cD$, and $c,s\in[0,1]$\\
\> Output: The list of irredundant association rules from $\cD$ with confidence $c$ and support $s$ \\\\
1.\> $\cR\Is\emptyset$\\
2.\>  $t\Is s|\cD|$, $\cG\Is$ GEN$_{\cI(\cF_{\cD,t})^-}(2^V,\cD,t)$\\
3. \> \For $i=n$ downto $1$, \Do\\
4.\>\> \Foreach $Z\in\cG$ with $|Z|=i$ \Do\\
5.\>\>\> $\cX(Z)\Is\Min\{X\in\cup_{j\not\in Z}\cX(Z\cup\{j\})~:~X\subseteq Z\}$\\
6.\>\>\> $t'\Is\frac{|S_{\cD}(Z)|}{c}+1$\\
7.\>\>\> $\cX(Z)\Is\cX(Z)\cup$ GEN-INC$_{\cF_{\cD[Z],t'}}(2^{Z},\cD[Z],t',\cX(Z),|\cF_{\cD[Z],t'}\setminus\cX(Z)|)$\\
8.\>\>\> $\cR\Is\cR \cup \{(X,Z): X\in\cX(Z)\setminus\bigcup_{j\not\in Z}\cX(Z\cup\{j\})\}$\\
9.\> \Return $\cR$\\
\end{code}
\caption{Generating irredundant association rules.}
\label{f-p1}
\end{figure}
  
\bigskip

We leave it as an exercise for the reader to verify that the procedure outputs all irredundant rules without repetition.

\medskip

The number of sets generated in the first step might be exponential in the number of irredundant rules. This is because some set $Z$ maybe 
frequent, but still there exist no \emph{new} minimal infrequent elements in $\cX(Z)$.
However, this seems unavoidable as the problem of generating the irredundant rules turns out to be NP-hard. To see why this is the case,
we note first that in \cite{BGKM03} it was proved that generating maximal frequent sets is hard.

\begin{theorem}\label{t1}{\em (\cite{BGKM03})}
Given a database $\cD\subseteq 2^{V}$ of binary attributes, and a
threshold $t$, problem INC-GEN$_{\cI(\cF_{\cD,t})}(2^V,\cD,t,\cX,1)$ is NP-hard.   
\end{theorem}

This immediately implies the following.

\begin{corollary}\label{c1}
Given a database $\cD\subseteq 2^{V}$ of binary attributes, and a
threshold $t$, the problem of generating all irredundant association rules is NP-hard. 
\end{corollary}
\proof
Consider the problem of generating maximal $t$-frequent sets.
Set $s=t/|\cD|$ and $c=1/|\cD|$. Then irredundant association rules are in one-to-one correspondence with 
minimal $X\subseteq V$ and maximal $Z\subseteq V$ satisfying \raf{eq1} and \raf{eq2}, and such that $X\subseteq Z$. 
By our choice of $c$ any such $X$ will be empty and thus the irredundant rules are in one-to-one correspondence with maximal sets $Z$ 
such that $|S(Z)|\geq t$. Thus Theorem \ref{t1}
implies that the problem of generating these rules is NP-hard.
\qed

Another framework to reduce redundancy, based on the concept of \emph{closed frequent} itemsets, is proposed in \cite{Z00}.

\subsubsection{Generalized association rules}\label{sgen}
We assume that each poset $\cP_i$ has a minimum element $l_i$. Following Definition \ref{d7}, we can generalize binary association  rules to more general databases as follows.
\begin{definition}\label{d8}\emph{(Irredundant generalized association rules)}
Let $\cD\subseteq \cP=\cP_1\times\cdots\times\cP_n$ be a database over a poset product, and $s,c\in[0,1]$ be given numbers. 
An irredundant association rule  $x\Rightarrow z |(c,s)$, with support $s$ and confidence $c$, 
is determined by a pair of a minimal element 
$x\in\cP$ and a maximal element $z\in\cP$, such that $x\preceq z$, $x_i\in\{z_i,l_i\}$ for all $i$, and
\begin{equation}\label{ardef}
\frac{|S_{\cD}(z)|}{|\cD|} \geq s,~~~~~~~\frac{|S_{\cD}(z)|}{|S_{\cD}(x)|}\geq c.
\end{equation}
\end{definition}

The rule $x\Rightarrow z$ is interpreted as follows: With support $s$, at least $c$ fraction of the transactions that dominate $x$ also dominate $z$ (i.e.,
$t\succeq x$ implies $t\succeq z$ for all $t\in \cD$). From the pair $(x,z)$, we can get a useful rule by letting $R=\{i~:~z_i=l_i\}$ and $S=\{i~:~x_i=z_i\}$, and inferring for a transaction $t\in\cD$ that 
\begin{equation}\label{eq3}
(t_i \succeq z_i) ~\forall i\in S\setminus R \Longrightarrow  (t_i \succeq z_i) ~\forall i\not\in S\cup R. 
\end{equation}

As in the binary case, the generation of such rules can be done, by first generating frequent elements from $\cD$ (working on a product of posets), then generating 
minimal frequent elements on a binary problem, defined by setting each $\cP_i=\{l_i,z_i\}$. In Appendix A, we give an extension of the Apriori Algorithm \cite{AS94} for
finding frequent elements in a database defined over a product of posets. 
 
As we shall see in the examples below, this generalization allows us to discover association rules in which antecedents and consequents 
are generalizations of the individual entries appearing in the database, and which might otherwise lack enough support.

\begin{example}\label{e4}(Association rules deriven from taxonomies)
Consider the database in Table \ref{tb3}. Using $s=0.3$ and $c=0.6$, we get $z=(\mbox{Outwear,Hiking Boots})$ as a frequent element, 
and $x=(\mbox{Outwear,Item})$ as a minimal infrequent element with $x\preceq z$ and $x\in\{\mbox{Item,Outwear}\}\times\{\mbox{Item,Hiking Boots}\}$. 
According to \raf{eq3}, this gives rise to the rule $\mbox{Outwear}\Rightarrow\mbox{Hiking Boots}$. Note that both rules $\mbox{Ski Pants}\Rightarrow\mbox{Hiking Boots}$ and
$\mbox{Jackets}\Rightarrow\mbox{Hiking Boots}$ lack minimum support, and hence the generalized association rule was useful.
\end{example}

In \cite{SA96}, a method was proposed for mining quantitative association
rules by partitioning the range of each quantitative attribute into
disjoint intervals, and thus reducing the problem into the
Boolean case. However, as mentioned in \cite{SA96}, this technique is
sensitive to the number of intervals selected for each attribute: if
the number of intervals is too small (respectively, too large), some
rules may not be discovered
since they lack minimum confidence (respectively, minimum support);
see \cite{SA96} for more details. 

\begin{figure}[t]
  \centerline{
\parbox{3in}{\psfig{figure=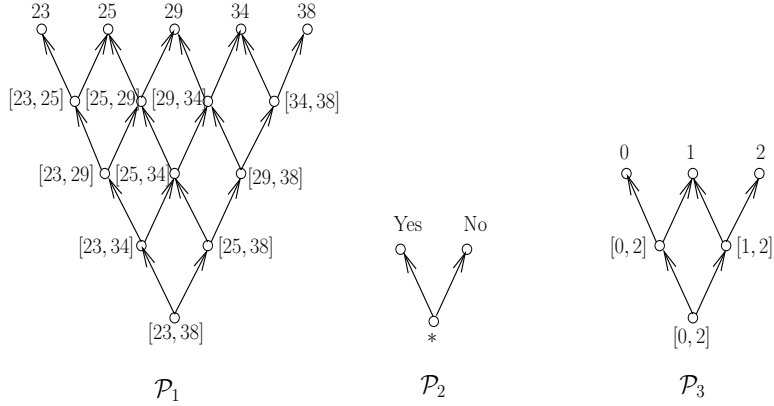,width=4.00in,height=2.1in}}}
\caption{The 3 factor posets in Example \ref{e5}.}
\label{f4}
\end{figure}

An alternative approach, which avoids the need to impose a certain
partitioning on the attribute ranges, is to consider each quantitative attribute as defined on a semi-lattice of intervals. 
More precisely, suppose that $a_i$ is a quantitative attribute, and consider the set of possible values
assumed by $a_i$ in the database, say, $\cS_i\defeq\{t_i~|~t\in\cD\}$.
Let $\cP_i$ be the \emph{dual} of the lattice of intervals whose elements correspond to the
different intervals defined by the points in $\cS_i$,
and ordered by containment. The minimum element $l_i$
of $\cP_i$ corresponds to the interval spanning all the points in $\cS_i$. The maximum element is not needed and can be deleted to
obtain a meet semi-lattice $\cP_i$. A $2$-dimensional example is shown in Figure \ref{f4}. 
Let $\cP=\cP_1\times\cdots\times\cP_n$. Then each element $x$ of $\cP$ corresponds to an $n$-dimensional box, and those elements can be used to 
produce association rules derived form the data. Using a similar reduction as the one that will be used in Section \ref{box}, 
the situation can be simplified since each semi-lattice $\cP_i$ can be decomposed into the product of two chains.

\medskip

For categorical attributes, each attribute value can be used to introduce a binary attribute. However, this imposes that each generated association rule
must have a condition on this attribute, which restricts the sets of rules generated. For example, in the database in Table \ref{tb2}, the categorical attribute "Married" 
can be replaced by two binary attributes "Married: Yes" and "Married: No" and an entry of "$1$" is entered in the right place in each record. But since each record must have a "$1$" in exactly one of these locations, this means that any association rule generated from this database must contain a condition on the marital status of the individual.  
Here is a way to avoid this restriction. For a categorical attribute $a_i$ which assumes values $\{v_1,\ldots,v_r\}$, we introduce an artificial element $l_i$ (corresponding essentially to a "don't care") and define a tree poset on $\{l_i,v_1,\ldots,v_r\}$ in which the only precedence relation are $l_i\prec v_j$, for $j=1,\ldots,r$ (see Figure \ref{f4}). 

Let us look at an example.

\hide{Clearly, each poset $\cP_i$ is a special case
forest, and therefore Theorems \ref{t0} and \ref{t1} apply here. In
fact, in the case of chains, it is easy (using binary search) to
modify the bounds on the running time in these theorems to make it only 
defendant on the {\em logarithm} of the range (i.e., the bit size) of
each attribute.}

\begin{example}\label{e5}(Quantitative association rules)
Consider the database in Table \ref{tb2}. This database can be viewed as a subset of the product of the $3$ posets shown in Figure \ref{f4}.
Using $s=0.4$ and $c=1.0$, we get $z=([34,38],\mbox{Yes},[2,2])$ as a frequent element, 
and $x=([34,38],*,[0:2])$ as a minimal infrequent element with $x\preceq z$ and $x\in\{[23,38],[30,38]\}\times\{*,Yes\}\times\{[0,2],[2,2]\}$ (assuming Age is integer-valued). 
According to \raf{e3}, this gives rise to the rule: $<\mbox{Age: } 34..38>\Rightarrow<\mbox{ Married: Yes}>\mbox{ and }<\mbox{NumCars: } 2>$. Note that the rule $(<\mbox{Age: } 34..38>\mbox{ and }<\mbox{ Married: Yes}>)\Rightarrow<\mbox{NumCars}: 2>$ is also valid but it is redundant since it is implied by the first rule.
\end{example}

Note that, using this approach, we consider overlapping two-sided
intervals for each attribute $a_i$, i.e., intervals of the form
$x_i\leq a_i\leq y_i$, but we do not set, a priori, the boundaries of
these intervals. Instead, these boundaries are determined by the
minimum support requirements and the values of the transactions in
the database. 

\hide{It should be also mentioned that this method may result
in more general association rules than the ones obtained by
partitioning ranges into intervals. Consider for example (see \cite{}) 
a database of $3$ attributes: $a_1\in\cP_1=\{20,\ldots,90\}$ , $a_2\in\cP_2=\{0,1\}$, and $a_3\in\cP_3=\{0,1,\ldots, 10\}$ with the data set
shown in Table \ref{table1}.

Using minimum support $s=0.4$ and minimum confidence $c=0.5$, we
obtain, as an example, the association rules: $(24,0,0)\Rightarrow
(34,1,2)$, $(20,1,0)\Rightarrow(29,0,0)$. The first rule can be
simplified to $a_1\geq 24 \Rightarrow a_2\geq 1, a_3\geq 2$, and
the second can be simplified to $20\leq a_1\leq 28 \Rightarrow a_2=0$,
etc.   
}

We refer the reader to \cite{HCC93,HF95,HMWG98,HW02,NCJK01,SA95,SA96,TS98,TYZ05} for more algorithms for mining generalized and quantitative association rules.
\subsubsection{Negative correlations}
Consider a binary database $\cD\subseteq 2^V$.
It may be interesting to generate association rules in which the antecedent or the consequent has a negated predicate. For instance, in Example \ref{e1}, 
we may be interested in generating also rules of the form: $(\mbox{Bread, Butter, Milk})\Rightarrow\neg \mbox{Yogurt}$, that is, customers who purchase
Bread, Butter, and Milk tend not to buy Yogurt.   

Several techniques have been proposed in the literature for mining negative correlations, see e.g. \cite{AZ04,BMS97,KP07,SVTV05,YBYZ02}.
Interestingly, such association rules can be found by embedding the database into the product of tree posets as follows. For each item we introduce a tree poset $\{*,+,-\}$,
where "$+$" stands for the item being present and "$-$" stands for the item being absent, and "$*$" stands for a "don't care". The only relations in this poset 
are $*\prec +$ and $*\prec -$. 

\begin{example}\label{e6}(Negative association rules)
Consider the database in Table \ref{tb1}. To allow for negative correlations, we view this database as a subset of the product $\cP$ of $6$ tree posets, as described above.
Using this representation, transaction $T_8$ in the table, for instance, corresponds to the element $x=(-,+,+,+,-,-)$ of $\cP$. Using $s=0.3$ and $c=0.75$, we get $z=(+,+,*,-,*,-)$ (corresponding to $\{\mbox{Bread, Butter, No Milk, No Yogurt}$) as a frequent element, and $x=(*,+,*,-,*,*)$ as a minimal infrequent element with $x\preceq z$. 
According to \raf{eq3}, this gives rise to the rule: $(\mbox{Butter,} \neg\mbox{Milk}) \Rightarrow(\mbox{Bread,}\neg\mbox{Yogurt})$.
\end{example}

\subsection{Generating rare associations and rare association rules}\label{rareAssoc}
In the examples we have seen above, our objective was to discover correlations that might exist between data attributes.  
In some situations, it maybe required to discover correlations in which some attributes are unlikely to assume certain values together.
This is a direct application of finding infrequent elements. Given a database $\cD\subseteq \cP_1\times\cdots\times\cP_n$, an infrequent element is a collection of 
generalizations of items that do not tend to appear together in the database. For instance, consider the database in Table \ref{tb3}. 
For $t=2$, the element $(\mbox{Jacket, Hiking Boots})$ is $t$-infrequent and we can conclude that in less than $34\%$ of the transactions these two items are purchased together. 
However, this is not the strongest conclusion we can make, since the minimal $t$-infrequent element $(\mbox{Jacket,Footwear})$ tells us that less than $34\%$ of the customers purchase jackets and footwear in a single transaction. 

\medskip

One important application of finding rare associations is in mining the so-called \emph{rare association rules}. These are rules that appear with \emph{low} support but \emph{high} confidence.
This happens when some of the items appear rarely in the database, but they exhibit enough association between them to generate useful rules.
The problem in discovering such rules is that one needs to set the minimum support parameter $s$ at a low value to be able to detect these rules, but this on the other hand, may introduce many other meaningless rules, resulting from other frequent itemsets, that would lack enough support otherwise\footnote{this dilemma is called the \emph{rate item problem} in \cite{M98}}. 
A number of methods have been proposed for dealing with such rare rules, see e.g. \cite{LHM99,K08}. One approach that can be used here is based on finding minimal infrequent elements.
Consider for simplicity a binary database $\cD\subseteq 2^V$. We choose two threshold values $0<s_1<s_2<1$ for the support: A subset of items $X\subseteq V$ will qualify if its support satisfies $s_1|\cD|\leq|S_{\cD}(X)|\leq s_2|\cD|$. Such sets will have enough support but still are infrequent.
Once these sets are generated, the discovery of the corresponding association rules can be done by looking at the confidence as before. 
The generation of these sets can be done as follows. First, we find the family $\cX$ of all minimal sets $X$ such that $|S_{\cD}(X)|\leq s_2|\cD|$, which is an instance of 
problem GEN$_{\cF_{\cD,t}}(2^V,\cD,t)$, with $t=s_2|\cD|$. Next, for each such $X\in\cX$, we find the frequent sets containing $X$, by solving an instance of problem GEN$_{\cI(\cF_{\cD',t'})^-}(2^V,\cD',t')$, where $\cD'=\{T\in \cD: T\supseteq X\}$ and $t'=s_1|\cD|$. A related approach was used in \cite{MNESM06}. 

\medskip

We look at two more examples of this kind in the next two subsections.  

\subsubsection{Maximal $k$-boxes} \label{box}
As another example\footnote{taken from \cite{EGLM01}}, consider a database of tickets, car registrations, and drivers' information.
Interesting observations that can be drawn from such tables could be: "No tickets were issued to BMW Z3 series cars before 1997", or "No tickets
for \$1000 were issued before 1990 for drivers born before 1956 ", etc.

To model these scenarios, we let $\cS$ be a set of points in $\RR^n$, representing the quantitative parts of the transactions in the database. 
We would like to find all regions in $\RR^n,$ which contain no, or a few, data points from $\cS$. Moreover, to avoid redundancy we are interested in finding
only maximal such regions. 
This motivates the following definition.

\begin{definition}\label{d9}\emph{(Maximal $k$-boxes)}
Let $\cS\subseteq\RR^n$ be a set of $n$-dimensional points and $k\leq |\cS|$ be a given integer. 
A {\em maximal $k$-box} is a closed $n$-dimensional box which contains at most $k$ points of $\cS$ in its interior, and which is maximal with respect
to this property (i.e., cannot be extended in any direction without strictly enclosing more points of $\cS$). 
\end{definition}

\begin{figure}[t]
  \centerline{
\parbox{2.75in}{\psfig{figure=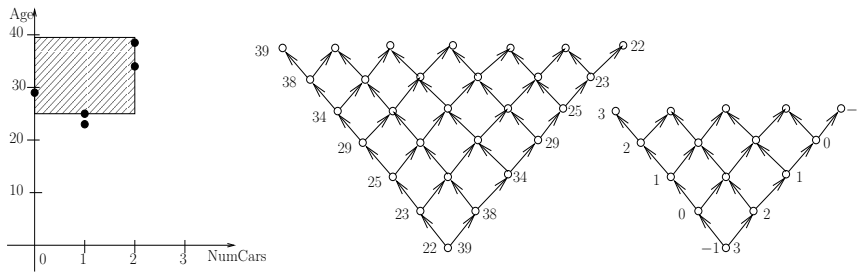,width=4.5in,height=1.4in}}}
        \hspace{1.0in}
\caption{A maximal empty box and the two factor posets used for representing such boxes.}
\label{f7}
\end{figure}

\begin{example}\label{e7}
Consider again the database in Table \ref{tb2}. In Figure \ref{f7}, we represent (Age,NumCars) as points in $2$-dimensional space.
The corresponding two products of chains are shown on the right. 
The box $B_1=[(25,0),(39,2)]$ is a maximal empty box, and box $B_2=[(23,0),(39,2)]$ is a maximal $1$-box. The box $B_1$ tells us that
no individuals with age between 26 and 38 have 1 car. 
\end{example}

\hide{\begin{table}[htbp]
\centering\small
\begin{tabular}{|c|c|c|c|c|c|c|}
\multicolumn{3}{c}{Registration} & \multicolumn{1}{c}{}&\multicolumn{3}{c}{Drivers} \\  \cline{1-3} \cline{5-7}
RegNum & Model & Owner & &  DLNum & Name & DOB\\ \cline{1-3} \cline{5-7}
R43999 & Saab95W & Owen & & G16999 & Smith & 1970-03-22 \\ \cline{1-3} \cline{5-7}
R44000 & HondaCW & Wang & & G65000 & Simon & 1908-03-05 \\ \cline{1-3} \cline{5-7}
$\cdots$ & $\cdots$ & $\cdots$ & & $\cdots$ & $\cdots$ & $\cdots$ \\ \cline{1-3} \cline{5-7}
\end{tabular}
\\
\caption{A database of registration and drivers}
\label{tb5}
\end{table}

\begin{table}[htbp]
\centering\small
\begin{tabular}{|c|c|c|c|c|c|c|}
\multicolumn{7}{c}{Tickets}\\\hline
$TID$ & Officer & RegNum & Date & Infraction & Amt & DLNum \\ \hline
119 & Seth & R43999 & 1/1/99 & Speed & 100 & G4337 \\ \hline
249 & Murray & R00222 & 2/2/95 & Parking & 30 & G7123 \\ \hline
$\cdots$ & $\cdots$ & $\cdots$ & $\cdots$ & $\cdots$ & $\cdots$ & $\cdots$ \\ \hline
\end{tabular}
\\
\caption{A database of tickets}
\label{tb6}
\end{table}
}

Let $\cF_{\cS,k}$ be the set of all maximal $k$-boxes for a given pointset $\cS$. Then we are interested in generating the elements of $\cF_{\cS,k}$.
Let us note that without any loss of generality, we could consider
the generation of the boxes $\{ B\cap D\mid B\in \cF_{\cS,k}\}$, where $D$ is a fixed bounded box containing all points
of $\cS$ in its interior. Let us further note that the $i$th coordinate of each vertex of such a box is the same as
$p_i$ for some $p\in \cS$, or the $i$th coordinate of a vertex of $D$, hence all these coordinates belong to a finite
set of cardinality at most $|\cS|+2$. Thus we can view $\cF_{\cS,k}$ as a set of boxes with vertices
belonging to such a finite grid.
More precisely, let $\cC_i=\{ p_i \mid
p\in \cS\}$ for $i=1,\ldots,n$ and consider the family of boxes $\cB=\{ [a,b]\subseteq \mathbb{R}^n
\mid a,b\in \cC_1 \times \cdots \times \cC_n,~ a\leq b\}$.
For $i=1,\ldots,n$, let $u_i=\max \cC_i$, and let
$\cC_i^*\defeq \{u_i-p\mid p\in \cC_i\}$ be the chain ordered in
the direction opposite to $\cC_i$.
Consider the $2n$-dimensional box $\cC=\cC_1\times \cdots \times \cC_n\times \cC_1^*\times \cdots \times
\cC_n^*$, and let us represent
every $n$-dimensional box $[a,b]\in\cB$ as the $2n$-dimensional vector $(a,u-b)\in \cC$, where $u=(u_1,\ldots,u_n)$.
This gives a monotone injective mapping $\cB\mapsto\cC$ (not all elements of $\cC$ define a box, since $a_i>b_i$ is possible
for $(a,u-b)\in\cC$).

It is not difficult to see that our problem reduces to solving problem GEN$_{\cF_{\cS,k+1}}(\cC^*,\cD,k+1)$, where $\cD\defeq\{(p,u-p)~:~p\in\cS\}$ and we redeine support to be $S_{\cD}(p)=\{q\in\cD~:~q\succ p\}$  (ignoring a small number (at most $\sum_{i=1}^n|\cC_i|$) of additionally generated elements, corresponding to non-boxes), see \cite{KBEGM07} for more details. 

\subsection{Minimal infrequent multi-dimensional intervals}\label{s-time-assoc}
Consider the database of intervals given in Section \ref{s-time}.  
An interesting observation, that may be deduced from the database, can take
the form ``Fewer than 40\% of the customers occupy the service on Friday between
2:00-3:00 and on Saturday between 2:00-4:00'', or "With support 60\%, all customers who make full use of the service between 2:00-3:00 on Friday 
tend also to use the service between 2:00-3:00 on Saturday and between 1:00-2:00 on Sunday". 
These examples illustrate the requirement for discovering correlations or association rules between occurrences of events over time. 
As in the previous examples, a fundamental problem that arises in this case is the generation of frequent
and minimal infrequent multi-dimensional intervals. 

More Formally, given a database of $n$-dimensional intervals $\cD$, and $i\in [n]$,
let $\mP_i=\{p_i^1,p_i^2,\ldots,p_i^{k_i}\}$ be the set of end-points of
intervals appearing in the 
$i$th column of $\cD$. Clearly $k_i\le 2|\cD|$, and assuming that
$p_i^1<p_i^2<\ldots<p_i^{k_i}$, we obtain a set
$\mI_i=\{[p_i^1,p_i^2],[p_i^2,p_i^3],\ldots,[p_i^{k_i-1},p_i^{k_i}]\}$
of at most $2|\cD|$ intervals. Let $\cP_i$ be the lattice of
intervals defined by the set $\mI_i$ (recall Definition \ref{d1}), for $i=1,\ldots, n$, and let
$\cP=\cP_1\times\cdots\times\cP_n$. Then, each record in $\cD$ appears as
an element in $\cP$, i.e., $\cD\subseteq\cP$. 
   
Now, it is easy to see that the $t$-frequent elements of $\cP$ are in one-to-one correspondence with
the $t$-frequent intervals defined by $\cD$, in the obvious
way: if $x=(x_1,\ldots,x_n)\in\cP$ is a frequent element,
then the corresponding interval $(I_1,\ldots,I_n)$ (where $I_i$
corresponds to $x_i$, for $i=1,\ldots,n$) is the corresponding
frequent interval.  The situation with minimal infrequent
intervals is just a bit more complicated: if
$x=(x_1,\ldots,x_n)\in\cP$ is a minimal infrequent element then the
corresponding minimal infrequent interval $(I_1,\ldots,I_n)$ is
computed as follows. For $i=1,\ldots,n$, if $x_i=l_i$ is the minimum
element of $\cP_i$, then $I_i=\emptyset$. If $x_i$ represents a point
$p_i\in\RR$ then $I_i=[p_i,p_i]$. Otherwise, let $[a_i,b_i]$ and
$[c_i,d_i]$ be the two intervals corresponding to the two immediate
predecessors of $x_i$ in $\cP_i$, where we assume $a_i<c_i$. If $a_i=b_i$ and
$c_i=d_i$ then $x_i$ corresponds to the interval $[a_i,c_i]$ 
and we have an infinite number of minimal infrequent
intervals defined (uniquely) by $I_i$, namely $I_i=[p_i,p_i]$ for all
points $p_i$ in the open interval $(a_i,c_i)$. Finally, if $a_i<b_i$
and $c_i<d_i$, then $I_i=[c_i-\epsilon,b_i+\epsilon]$ for a
sufficiently small constant $\epsilon$ (which can be taken as the smallest precision used in the representation of intervals, e.g., 1 minute).     
Consequently, in all cases, our problems reduce to finding
$t$-frequent/minimal $t$-infrequent elements in the lattice product $\cP$.

\section{Complexity}\label{s5}
\subsection{Minimal infrequent elements}
We will illustrate now that, for all the examples considered above, the problem of finding minimal $t$-infrequent elements,
that is, problem GEN$_{\cF_{\cD,t}}(\cP,\cD,t,\cX)$ can be solved in incremental quasi-polynomial time.

Central to this is the notion of \emph{duality testing}. Call two subsets $\cA,\cB\subseteq\cP$ {\it partially dual} if the following condition holds:
\begin{equation}\label{dual}
a\not\preceq b,~~\mbox{for all }a\in\cA, b\in\cB.
\end{equation}
For instance if $\cX\subseteq\cF_{\cD,t}$ and $\cY\subseteq\cI(\cF_{\cD,t})$ then $\cX,\cY$ are partially dual.
The duality testing problem on $\cP$ is the following:
\begin{description}
\item[DUAL$(\cP,\cA,\cB)$:] Given two partially dual sets $\cA,\cB\subseteq\cP$, check if there exists an element $x\in\cP$, such that 
\begin{equation}\label{cond1}
\mbox{$x\not\succeq a$ for all $a\in\cA$ and $x\not\preceq b$ for all $b\in\cB$. }
\end{equation}
\end{description} 

Let $m=|\cA|+|\cB|$. The main result that we need here is the following.
\begin{theorem}\label{t2}(\cite{BEGKM02-SICOMP,E08}) 

(i) If each $\cP_i$ is a chain, then DUAL$(\cP,\cA,\cB)$ can be solved in $n\cdot m^{o(\log m)}$ time.

(ii) If each $\cP_i$ is tree poset, then DUAL$(\cP,\cA,\cB)$ can be solved in $\poly(n,\mu(\cP))\cdot m^{o(\log m)}$ time, where $\mu(P)=\max\{|\cP_i|~:~i\in[n]\}$.

(iii) If each poset $\cP_i$ is a lattice of intervals then DUAL$(\cP,\cA,\cB)$ can be solved in $k^{O(\log^2 k)}$ time, where
$k=m+\sum_{i=1}^n|\cP_i|$.

\end{theorem}

We also note that a mixture of posets of the three types can be taken in the product and the running time will be the maximum of the bounds in (i), (ii) and (iii).
Thus the duality testing problem can be solved in quasi-polynomial time for the classes of posets that arise in our applications. 
To apply this result to the generation of minimal infrequent elements, we need another important ingredient. Namely, that the number of all \emph{maximal} $t$-frequent elements
is polynomially small in the number of minimal $t$-infrequent elements. In fact the following stronger bound holds.

\begin{theorem}[\cite{BGKM02}]\label{tbound}
For any poset product $\cP=\cP_1\times\ldots\times\cP_n$ in which each two elements of each poset $\cP_i$ have at most one join, 
the set $\cF_{\cD,t}$ is \emph{uniformly dual-bounded} in the sense that
\begin{equation}\label{ebound}
\left|\cI(\cA)\cap\cI(\cF_{\cD,t})\right|\leq (|\cD|-t+1)|\cA|,
\end{equation}
for any non-empty subset $\cA\subseteq \cF_{\cD,t}$.   
\end{theorem}

\medskip

To generate the elements of $\cF_{\cD,t}$ we keep two lists $\cX\subseteq\cF_{\cD,t}$ and $\cY\subseteq\cI(\cF_{\cD,t})$, both initially empty. 
Given these partial lists, we call the procedure for solving DUAL$(\cP,\cX,\cY)$. If it returns an element $x$ satisfying \raf{cond1}, we obtain from $x$ a vector $x'$ in $\cF_{\cD,t}$ or $\cI(\cF_{\cD,t})$, depending respectively on whether
$x$ is $t$-infrequent or $t$-frequent element. This continues until no more such elements $x$ can be returned. Clearly, if this happens then all elements of $\cP$ have been classified to either lie above some $x\in\cX$ or below some $x\in\cY$, i.e., $\cX=\cF_{\cD,t}$ and $\cY=\cI(\cF_{\cD,t})$. By \raf{ebound}, the time needed to produce a new element of $\cF_{\cD,t}$ is at most a factor of $|\cD|$ times the time needed to solve problem DUAL$(\cP,\cX,\cY)$. A Pseudo-code is shown in Figure \ref{f8}.
\begin{figure}
\begin{code}
\noindent{\bf Procedure GenerateInfrequent$(\cP,\cD,t)$:}\\
\>{\it Input:} A database $\cD\subseteq\cP$ and a integer threshold $t$.\\
\> {\it Output:} The $t$-minimal infrequent elements. \\\\
1.\> $\cX\Is\emptyset$; $\cY\Is\emptyset$.\\
2.\> \While DUAL$(\cP,\cX,\cY)$ returns a vector $x$\\
3.\>\> If $|S_{\cD}(x)|<t$, \Then\\
4.\>\>\>   $x'\Is$ a minimal vector such that $x'\preceq x$ and $|S_{\cD}(x)|<t$.\\
5.\>\>\>   $\cX\Is\cX\cup\{x'\}$.\\
6.\>\> \Else \\
7.\>\>\>   $x'\Is$ a maximal vector such that $x\preceq x'$ and $|S_{\cD}(x)|\ge t$ \\
8.\>\>\>   $\cY\Is\cY\cup\{x'\}$.\\
9.\> \Return $\cX$.\\
\end{code}
\caption{A procedure for enumerating minimal infrequent elements.}
\label{f8}
\end{figure}

\medskip

\begin{theorem}\label{t4}
Let $\cP=\cP_1\times\cdot\times\cP_n,$  where each $\cP_i$ is either a chain, a lattice of intervals, or a meet semi-lattice tree poset. Then for any $\cD\subseteq\cP$, and integer $t$,
problem GEN$_{\cF_{\cD,t}}(\cP,\cD,t)$ can be solved in incremental
quasi-polynomial time.
\end{theorem}

\medskip

In Appendix B, we give the dualization algorithm for meet semi-lattice tree posets. We refer the reader to \cite{E08} for more details and for the dualization algorithm on products of lattices of intervals.

\subsection{Infrequent/frequent elements}
If we are interested in finding all infrequent elements rather then the minimal ones, then the problem seems to be easier. 
As we have seen in the applications above, one basic step in finding association rules is enumerating all frequent elements. Those can be typically found by 
an Apriori-like algorithm, which we give for completeness in Appendix A. Since one can regard the problem of finding infrequent elements as of that finding frequent elements 
on the dual poset, we can conclude that the infrequent elements can also be found by the algorithm Apriori, and hence the problem can be solved in
incremental polynomial time. 

\begin{theorem}\label{t5}
Let $\cP=\cP_1\times\cdots\times\cP_n,$, $\cD\subseteq\cP$, and $t$ be an integer.
Then all $t$-frequent ($t$-infrequent) elements can be computed with amortized delay. 
\end{theorem}

\section{Conclusion}\label{s6}
In this chapter, we have looked at a general framework that allows us to mine associations from different types of databases. 
We have argued that the rules obtained under this framework are generally stronger than the ones obtained from techniques that use binarization.
A fundamental problem that comes out from this framework is that of finding minimal infrequent elements in a given product of partially ordered sets.
As we have seen, this problem can be solved in quasi-polynomial time, while the problem becomes easier if we are interested in finding all infrequent/frequent elements.
On the theoretical level, while the complexity of enumerating minimal infrequent elements is not known to be polynomial, the problem is unlikely to be NP-hard unless every NP-complete problem can be solved in quasi-polynomial time.

Finally, we mention that a number of implementations exist for the duality testing problem on products of chains \cite{BMR03,KS05,KBEG06}, and for the generation of infrequent elements \cite{KBEG06} on such products. Experiments in \cite{KBEG06} indicate that the algorithms behave practically much faster than the theoretically best-known upper bounds on their running times, and therefore may be applicable in practical applications. Improving these implementations further and putting them into practical use, as well as the extension to more general products of partially ordered sets remain challenging issues that can be the subject of interesting future research.

\section*{Appendix A: Frequent elements generation - Apriori algorithm}\label{a}
Let $\cP=\cP_1\times\cdots\times\cP_n$ be a product of $n$ posets and $\cD\subseteq\cP$ be a database. For simplicity,
we assume that $\cP$ has a minimum element $l=(l_1,\ldots,l_n)$.
Given an integer threshold $t$, we present below
an Apriori-like algorithm that finds all the $t$-frequent elements $x\in\cP$.
This can be viewed as a strict generalization of the one for frequent itemsets in \cite{AS94}.
The algorithm for generating all infrequent elements is exactly the
same, but it should work on the dual poset $\cP^*$. 
We assume that, for $i=1,\ldots,n$, each element in
$x\in\cL_i$ is assigned a number $d(x)$ that indicates the {\em
longest} distance, in the precedence graph of $\cP_i$, from the
smallest element 
$l_i$ of $\cL_i$ to $x$ (such numbers are easy to compute since the
precedence graph is acyclic). For $x=(x_1,\ldots,x_n)\in\cP$, we let
$d(x)=\sum_{i=1}^n d(x_i)$. We say that $x$ has level $k$ is $d(x)=k$.

For $x\in\cP_i$, denote by $x^{\bot}$ the set of immediate
predecessors of $x$, i.e., $$x^\bot=\{y \in \cP_i~|~y \prec  x,~(\nexists z\in\cP_i :
y\prec z \prec x)\}.$$  
Similarly, denote  by  $x^{\top}$ the set of immediate successors of
$x$.
Note that, 
given $x=(x_1,\ldots,x_n)\in\cP$, the immediate predecessors of $x$ are 
given by:
$x^{\bot}=\{y\in\cP~|~y_i\in x_i^{\bot}\mbox{ for some }i\in [n]\mbox
{ and } x_j=y_j \mbox{ for all }j\neq i\},$
and let $\indeg(\cP)= \max\{|x^\bot|:~x \in \cP \}$.
The immediate successors of $x$ are similarly defined, and we let
$\outdeg(\cP)= \max\{|x^\top|:~x \in \cP \}$. 
Thus
$|x^\bot|=\sum_{i=1}^n|x_i^\bot|$ and
$|x^\top|=\sum_{i=1}^n|x_i^\top|$, for any $x=(x_1,\ldots,x_n)\in\cP$.

As in the standard Apriori algorithm for finding frequent sets, the levelwise procedure proceeds bottom-up on the levels of the poset 
preforming two basic steps at each level $k$: Candidate generation and pruning. In the first step, we generate a set of $\cC$ of candidate frequent elements
at level $k$, based on the set $\cF_{k-1}$ of frequent elements that we have already produced level $k-1$. In the pruning step, this set of candidates is
scanned keeping only the set if frequent elements. The procedures are shown in Figures \ref{Ap-f1}-\ref{Ap-f3}. 
  
\medskip

\begin{figure}
\begin{code}
\noindent{\bf Procedure Ariori$(\cP,\cD,t)$:}\\
\>{\it Input:} A database $\cD\subseteq\cP$ and a integer threshold $t$.\\
\> {\it Output:} The $t$-frequent elements. \\\\
1.\> $k\leftarrow 0$; $\cF_k\leftarrow\{l\}$;\\
2.\> while $\cF_k\neq\emptyset$ do \\
3.\>\> $\cC\leftarrow$ Candidates($\cF_k,k$);\\
4.\>\> $\cF_{k+1}\leftarrow$ Prune($\cC,\cD,t$); \\
5.\>\> $k\leftarrow k+1$;\\
6.\> end\\
7.\> return $\bigcup_{j=1}^k\cF_j$;
\end{code}
\caption{A procedure for enumerating frequent elements.}
\label{Ap-f1}
\end{figure}

\medskip

\begin{figure}
\begin{code}
\noindent{\bf Procedure Candidates$(\cP,\cF_k,k)$:}\\
\>{\it Input:} A poset $\cP$, an integer $k$ and a set of frequent elements at level $k$.\\
\> {\it Output:} A set of candidate frequent elements at level $k+1$. \\\\
1.\> $\cC\leftarrow\emptyset$;\\
2.\> for all $x\in \cF_k$ do\\
3.\>\> for all $y\in x^{\top}$ such that $d(y)=k+1$ do \\
4.\>\>\>if $\forall z\in y^{\bot}$ such that $d(z)=k$: $z\in \cF_k$, then \\
5.\>\>\>\> $\cC\leftarrow \cC\cup\{y\}$;\\
6.\> return $\cC$;
\end{code}
\caption{A procedure for level $(k+1)$-candidate generation.}
\label{Ap-f2}
\end{figure}

\medskip

\begin{figure}
\begin{code}
\noindent{\bf Procedure Prune$(\cC,\cD,t)$:}\\
\>{\it Input:} A database $\cD\subseteq\cP$, a integer threshold $t$, and a set of level $k$-candidates.\\
\> {\it Output:} The $t$-frequent elements among $\cC$. \\\\
1. $\cF\leftarrow\emptyset$;\\
2.\> for all $x\in \cC$ do\\
3.\>\> if $|S_{\cD}(x)|\geq t$ then \\
4.\>\>\> $\cF\leftarrow \cF\cup\{x\}$;\\
5.\> return $\cF$;
\end{code}
\caption{A procedure for extracting frequent elements from candidates.}
\label{Ap-f3}
\end{figure}

\medskip

Clearly, the number of scans of the database can be reduced by computing
the contribution of each transaction to the counts of all candidates
before reading the next transaction, see e.g. \cite{AS94}.      

Let $\tau$ be the maximum time required by the procedure to compute
the value of the function $|S_{\cD}(x)|$ for any $x\in\cP$.
\begin{lemma}\label{la1}
Algorithm Apriori outputs all $t$-frequent elements of $\cP$, with amortized delay $O(\indeg(\cP)\outdeg(\cP)(
n\sum_{i=0}^n\log|\cP_i|+\tau))$.
\end{lemma}
\proof
Let us note by induction on $k=0,1,\ldots$, that
$\cF_k=\cF_k'\defeq\{x\in\cP~:~d(x)=k, |S_{\cD}(x)|\geq t\}$.  Indeed, this
holds initially for $k=0$. Assume that it also holds for any $k>0$,
and consider the set $\cF_{k+1}$ generated in Step 4 of procedure
Apriori$(\cP,\cD,t)$. From Steps 3 in procedure Candidates$(\cP,\cF_k,k)$ and 3
in Prune$(\cC,\cD,t)$, we note that $\cF_{k+1}\subseteq\cF_{k+1}'$. So it
remains to show that $\cF_{k+1}'\subseteq\cF_{k+1}$. For this consider any
$y\in\cF_{k+1}'$ and observe, by the anti-monotonicity of $|S_{\cD}(\cdot)|$ and the
definition of $d(\cdot)$, that there exists an $x\in\cF_{k}'=\cF_k$ such
that $y\in x^{\top}$. Thus $x$ and $y$ pass respectively the tests in
Steps 2 and 3 of procedure Candidates$(\cF_k,k)$. Moreover, every
$z\in y^{\bot}$ with $d(z)=k$ belongs to $\cF_k'$ and hence to $\cF_k$
and therefore $y$ will be added to the list of candidates $\cC$ in
procedure Candidates$(\cF_k,k)$ and to the frontier list $\cF_{k+1}$
in Step 4 of procedure Prune$(\cC,\cD,t)$.    

Now we consider the running time of the procedure. Let
$k_{max}=\max\{k\in \ZZ_+~|~\cF_k\neq \emptyset\}$. By implementing a
balanced binary search tree on the elements of $\cF_k$ (sorted
according to some lexicographic ordering), we can perform the check
$z\in\cF_k$, for any $z\in\cP$, in $O(n\log|\cF_k|)$ time. Thus it
follows that the total time 
required by the procedure to output the union
$\cF_1\cup\ldots\cup\cF_{k_{max}}$ is bounded by
$$\sum_{k=0}^{k_{max}}\left(\sum_{x\in\cF_k}\sum_{y\in
x^{\top}}\left(\sum_{z\in y^{\bot}} O(n\log|\cF_k|)
+\tau\right)\right)
\leq \outdeg(\cP)\indeg(\cP)\sum_{k=0}^{k_{max}}|\cF_k|(O(n\log
|\cF_k|)+\tau). $$
This amounts to an amortized time of
$$
\indeg(\cP)\outdeg(\cP)\frac{\sum_{k=0}^{k_{max}}|\cF_k|(O(n\log
|\cF_k|)+\tau)}{\sum_{k=0}^{k_{max}}|\cF_k|}=
\indeg(\cP)\outdeg(\cP) O(n\log(\sum_{k=0}^{k_{max}}|\cF_k|)+\tau).
$$ 
Note that $\sum_{k=0}^{k_{max}}|\cF_k|\leq |\cP|=\prod_{i=1}^n|\cP_i|$,
and the lemma follows.
\qed

\section*{Appendix B: Dualization in products of meet semi-lattice tree posets}\label{trees}

Let $\cP=\cP_1\times\cdots\times\cP_n$, where the precedence graph of each poset $\cP_i$ is a meet semi-lattice tree poset (henceforth abbreviated MSTP),
and let $\cA,\cB\subseteq \cP$ two antichains satisfying \raf{dual}. We say that $\cB$ is {\em dual to}
$\cA$ if $\cB=\cI(\cA)$.

Note that in this case, we have the following decomposition of $\cP$
\begin{equation}\label{eb*}
\cA^+ \cap \cB^{-} = \emptyset,~~~~\cA^+ \cup \cB^{-}=\cP,
\end{equation}
and thus problem DUAL$(\cP,\cA,\cB)$ can be equivalently stated as follows:
\begin{description}
\item[{\em DUAL}$(\cP,\cA,\cB)$:] {\em Given antichains $\cA,\cB \subseteq
\cP$ satisfying \raf{dual}, check if there an $x\in\cP\setminus(\cA^+\cup\cB^-)$.}
\end{description}

Given any $\cQ\subseteq \cP$, let us denote by
$$\cA(\cQ)=\{a\in\cA~|~a^+\cap\cQ\neq\emptyset\},~~~~~~~
\cB(\cQ)=\{b\in\cB~|~b^-\cap\cQ\neq\emptyset\}.$$
These are the effective subsets of $\cA,\cB$ that play a role in problem DUAL$(\cQ,\cA,\cB)$.
Note that, for $a\in\cA$ and $\cQ=\cQ_1\times\ldots\times\cQ_n$, 
$a^+\cap\cQ\neq\emptyset$ if and only if $a_i^+\cap\cQ_i\neq\emptyset$,
for all $i\in[n]$. Thus, the sets $\cA(\cQ)$ and $\cB(\cQ)$ can
be found in $O(nm\mu(\cP))$ time.

To solve problem DUAL$(\cP,\cA,\cB)$,
we decompose it into a number of smaller subproblems which are solved
recursively. In each such subproblem, we start
with a subposet $\cQ=\cQ_1\times\ldots\times\cQ_n\subseteq\cP$
(initially $\cQ=\cP$), and two subsets
$\cA(\cQ)\subseteq\cA$ and $\cB(\cQ)\subseteq \cB$, and we want to check
whether $\cA(\cQ)$ and $\cB(\cQ)$ are dual in $\cQ$. 
The decomposition of $\cQ$ is done by decomposing one factor poset, say $\cQ_i,$ into a number of (not necessarily disjoint) 
subposets $\cQ_i^1,\ldots,\cQ_i^r,$ and solving $r$ subproblems on the $r$ different posets 
$\cQ_1\times\cdots\times\cQ_{i-1}\times\cQ_i^j\times\cQ_{i+1}\times\cdots\times\cQ_n,$ $j=1,\ldots,r$. 
For brevity, let us denote by
$\overline{\cQ}$ the product $\cQ_1\times\cdots\times\cQ_{i-1}\times\cQ_{i+1}\times\cdots\times\cQ_n,$, and
accordingly by $\overline{q}$ the element $(q_1,\ldots,q_{i-1},q_{i+1},\ldots,q_n)$,
for an element $q=(q_1,q_2,\ldots,q_n)\in\cQ$.

The algorithm is shown in Figure \ref{f7}. 
We assume that procedure $\fdtb$ returns either true or false depending on whether $\cA$ and $\cB$ are dual in $\cQ$ or not. Returning an element $x\in\cQ\setminus(\cA^+\cup\cB^-)$ in the latter case is straightforward, as it can be obtained from any subproblem that failed the test for duality. 

Note that after decomposing one of the posets, some elements $x\in\cA\cup\cB$ do not belong
to the current poset $\cQ$. In step 1, the elements that do not affect the solution are deleted, while in step 2, those that affect the solution
are projected down to the current poset $\cQ$ (by replacing each $a\in\cA$ ($b\in \cB$) with \emph{unique} element above $a$ (respectively, below $b$) in $\cQ$.)
In step 3, we check if the size of the problem is sufficiently small, and if so we use an exhaustive search procedure to decide the duality of $\cA$ and $\cB$ in $\cQ$. 

Starting from step 5, we decompose $\cQ\subseteq\cP$ by picking $a \in \cA$, $b \in
\cB$ and an $i \in [n]$, such that $a_i\not \preceq b_i$. The algorithm uses the effective volume $v=v(\cA,\cB)$ to compute the threshold 
$$
\epsilon(v)=\frac{1}{\chi(v)},~~~~\mbox{ where}~ \chi(v)^{\chi(v)}=v\defeq|\cA||\cB|.$$
If the minimum of $\epsilon^{\cA}\defeq|\cA_{\succeq}(a_i)|/|\cA|$ and $\epsilon^{\cB}\defeq|\cB_{\not\succeq}(a_i)|/|\cB|$,
where $\cA_{\succeq}(a_i)\defeq\{x\in\cA:x_i\succeq a_i\}$ and $\cB_{\not\succeq}(a_i)\defeq\{x\in\cB:x_i\not\succeq a_i\}$, 
is bigger than $\epsilon(v)$, then we decompose $\cQ_i$ into two MSTP's $\cQ_i'\leftarrow \cQ_i\cap a_i^+$ and $\cQ_i''\leftarrow \cQ_i\setminus\cQ_i',$ and solve
recursively two problems on these posets (steps 8 and 9). 

\begin{figure}
  \centerline{
\parbox{3in}{\psfig{figure=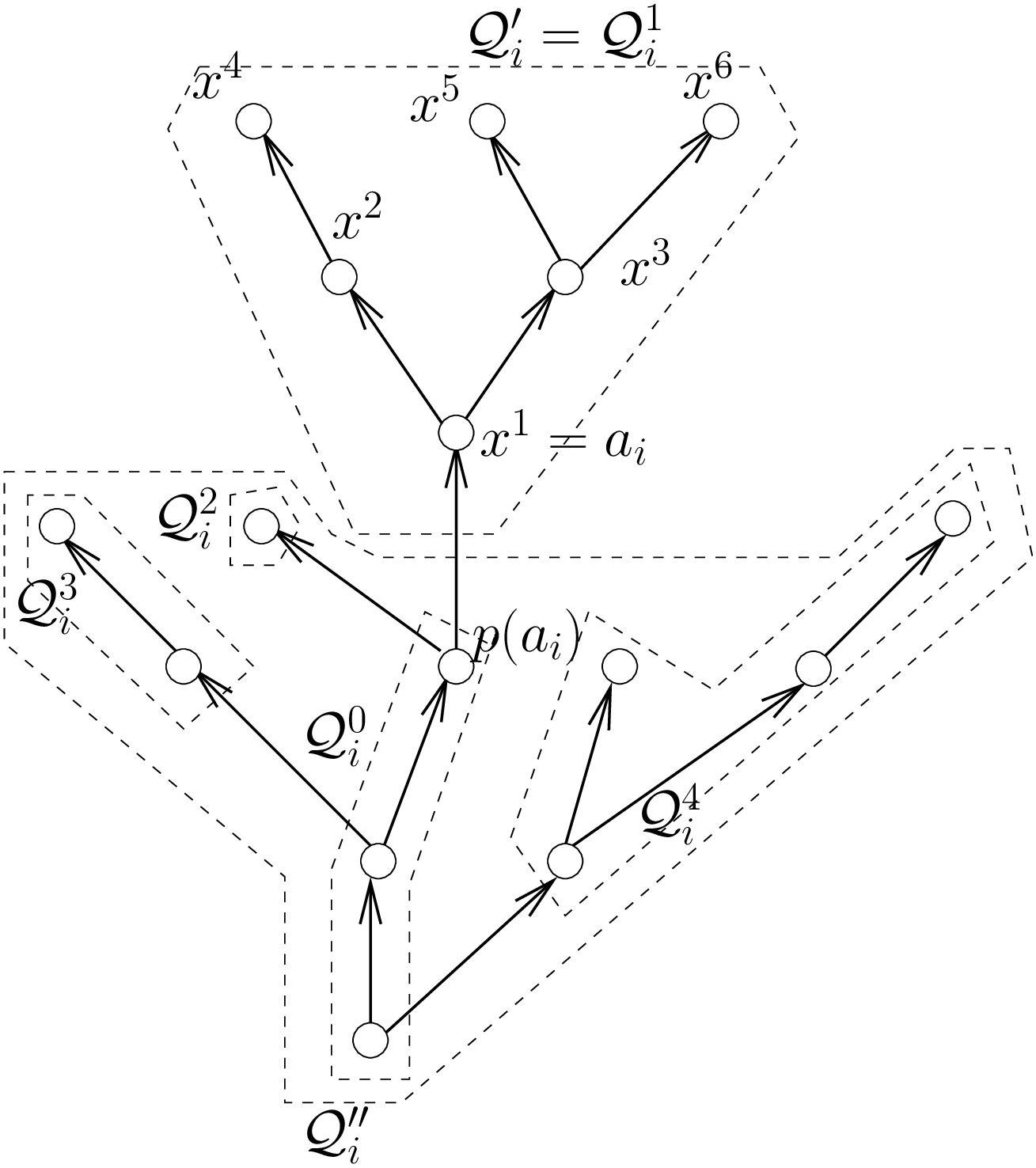,width=2.65in,height=2.5in}}}
\vspace{0.15in}
\caption{Decomposing the forest $\cQ_i$.}
\label{f11}
\end{figure}

Otherwise we proceed as follows. For $x\in\cQ_i$ denote by $p(x)$ the unique predecessor of $x$ in $\cQ_i$. 
Let $\cQ_i^0=p(a_i)^-\cap\cQ_i''$, $\cQ_i^1=\cQ_i',$ and $\cQ_i^2,\ldots,\cQ_i^r$ be the MSTP's obtained by deleting $p(a_i)^-$ from $\cQ_i''$ (see Figure~\ref{f11}).  
Then we can use the decomposition in step 12.

Finally, if $\epsilon^{\cA}\leq\epsilon(v)<\epsilon^{\cB}$, we proceed as in steps 14-17: we solve the subproblem on $\cQ_i''\times\overline{\cQ}$, and if
it does not have a solution $x$, then we process the elements $x^1,\ldots,x^k$
of $\cQ_i'$ in \emph{topological} order (that is, $x^j\prec x^r$ implies $j<r$). For each such element, we solve at most $|\cB|$ subproblems on 
$\{x^j\}\times(\overline{\cQ}\cap\overline{b}^-)$, for $b\in\cB_{\succeq}(p(x^j))$.

\bigskip

\begin{figure}
\begin{code}
\noindent{\bf Procedure $\fdtb(\cQ,\cA,\cB)$:}\\
\> Input: A subposet of a product of trees $\cQ=\cQ_1\times\cdots\times\cQ_n\subseteq\cP$ and two anti-chains $\cA,\cB\subseteq \cP$\\
\> Output: \True if $\cA$ and $\cB$ are dual in $\cQ$  and \False otherwise\\\\

1.\> $\cA\leftarrow\cA(\cQ)$, $\cB\leftarrow\cB(\cQ)$\\
2.\> $\cA\leftarrow\pr(\cQ,\cA)$, $\cB\leftarrow\pr(\cQ,\cB)$\\
3. \>\If $\min\{|\cA|,|\cB|\}\le 3$  \Then\\
4.\>\> \Return $\pd(\cQ,\cA,\cB)$\\
5.\> Let $a \in \cA$, $b\in \cB$, and $i \in [n]$ be such that $a_i\not \preceq b_i$\\ 
6.\> $\epsilon^{\cA}\leftarrow\frac{|\cA_{\succeq}(a_i)|}{|\cA|}$ and $\epsilon^{\cB}\leftarrow\frac{|\cB_{\not\succeq}(a_i)|}{|\cB|}$\\
7.\> Let $\cQ_i'\leftarrow \cQ_i\cap a_i^+,$ $\cQ_i''\leftarrow \cQ_i\setminus\cQ_i'$\\
8.\> \If $\min\{\epsilon^{\cA},\epsilon^{\cB}\}>\epsilon(v(\cA,\cB))$ \Then\\
9.\> \> \Return $\fdtb(\cQ_i'\times\overline{\cQ},\cA,\cB)\wedge \fdtb(\cQ_i''\times\overline{\cQ},\cA,\cB)$\\
10.\> \If $\epsilon^{\cB}\le\epsilon(v(\cA,\cB))$ \Then\\
11.\>\> Let $\cQ_i^0=p(a_i)^-\cap\cQ_i''$, $\cQ_i^1,\ldots,\cQ_i^r$ be the MSTP's composing $\cQ_i\setminus\cQ_i^0$  \\
12.\> \> \Return $\bigwedge_{j=1}^r \fdtb(\cQ_i^j\times\overline{\cQ},\cA,\cB)\wedge(\bigwedge_{a\in\cA_{\preceq}(a_i)}\fdtb(\cQ_i^0\times(\overline{\cQ}\cap\overline{a}^+),\cA,\cB))$\\ 
13.\> \Else\\
14.\> \> Let $x^1,\ldots,x^k$ be the elements of $\cQ_i'$ in topologically non-decreasing order\\
15.\> \> $d\leftarrow \fdtb(\cQ_i''\times\overline{\cQ},\cA,\cB)$\\
16.\> \> \For $i=1,\ldots,k$ \Do\\
17.\>\>\> $d\leftarrow d\wedge(\bigwedge_{b\in\cB_{\succeq}(p(x^j))}\fdtb(\{x^j\}\times(\overline{\cQ}\cap\overline{b}^-),\cA,\cB))$\\
18.\> \> \Return $d$ \\
\end{code}
\caption{The dualization procedure for MSTP's.}
\label{f9}
\end{figure}

\bibliographystyle{alpha}


\end{document}

%% file: makros2.tex
\newcommand{\realrange}[2]{\left[#1, #2\right]}

\newcommand{\unitrange}[2]{\realrange{0}{1}}

\newcommand{\llabel}[1]{\label{\labelprefix:#1}}
\newcommand{\labelprefix}{} 

\newcommand{\discussionsize}{\small}

\newenvironment{code}{\noindent
\begin{tabbing}%
\hspace{2em}\=\hspace{2em}\=\hspace{2em}\=\hspace{2em}\=\hspace{2em}\=%
\hspace{2em}\=\hspace{2em}\=\hspace{2em}\=\hspace{2em}\=\hspace{2em}\=%
\kill}{\end{tabbing}}

\newcommand{\labelcommand}{}
\newcommand{\captiontext}{}
\newsavebox{\codeparam}
\newcounter{lineNumber}
\newenvironment{disscodepos}[3]{%
\renewcommand{\labelcommand}{#2}%
\renewcommand{\captiontext}{#3}%
\sbox{\codeparam}{\parbox{\textwidth}{#3}}%
\begin{figure}[#1]\begin{center}\begin{code}\setcounter{lineNumber}{1}}{%
\end{code}\end{center}\caption{\llabel{\labelcommand}\captiontext}\end{figure}}

{\end{disscodepos}}

\newcommand{\While}    {{\bf while\ }}

\newcommand{\Do}       {{\bf do\ }}

\newcommand{\For}      {{\bf for\ }}
\newcommand{\Foreach}      {{\bf foreach\ }}
\newcommand{\Is}{\mbox{\rm := }}

\newcommand{\If}       {{\bf if\ }}

\newcommand{\Then}     {{\bf then\ }}
\newcommand{\Else}     {{\bf else\ }}

\newcommand{\Return}   {{\bf return\ }}

\newcommand{\True}     {{\bf true\ }}
\newcommand{\False}    {{\bf false\ }}

\newdimen\endofsize\endofsize=0.5em
\def\endofbeweis{~\quad\hglue\hsize minus\hsize
                 \hbox{\vrule height \endofsize width
\endofsize}\par}

